\documentclass[sigconf]{acmart}

\usepackage{amsmath}
\usepackage{algorithm}
\usepackage[noend]{algpseudocode}
\usepackage{graphicx}
\usepackage{textcomp}
\usepackage{xcolor}
\usepackage{hyperref}
\usepackage{subcaption}
\usepackage{makecell}
\def\BibTeX{{\rm B\kern-.05em{\sc i\kern-.025em b}\kern-.08em
    T\kern-.1667em\lower.7ex\hbox{E}\kern-.125emX}}

\newcommand{\sm}[1]{{\scriptstyle#1}}
\newcommand{\vsm}[1]{{\scriptscriptstyle#1}}

\renewcommand\footnotetextcopyrightpermission[1]{}

\begin{document}

\title{A Dynamic, Hierarchical Resource Model for Converged Computing}

\setcopyright{none}
\copyrightyear{2021}
\acmYear{2021}
\setcopyright{acmcopyright}
\acmConference[XXXX'21]{}{USA}

\author{Daniel J. Milroy}
\affiliation{%
  \institution{Lawrence Livermore National Laboratory}
  \city{Livermore}
  \state{CA}
  \country{USA}
}
\email{milroy1@llnl.gov}

\author{Claudia Misale}
\affiliation{%
  \institution{IBM T. J. Watson Research Center}
  \city{Yorktown Heights}
  \state{NY}
  \country{USA}
}
\email{c.misale@ibm.com}

\author{Stephen Herbein}
\affiliation{%
  \institution{Lawrence Livermore National Laboratory}
  \city{Livermore}
  \state{CA}
  \country{USA}
}
\email{herbein1@llnl.gov}

\author{Dong H. Ahn}
\affiliation{%
  \institution{Lawrence Livermore National Laboratory}
  \city{Livermore}
  \state{CA}
  \country{USA}
}
\email{ahn1@llnl.gov}

\begin{abstract}
{\em Extreme dynamic heterogeneity} in high performance computing systems
and the convergence of traditional HPC
with new simulation, analysis, and data science approaches
impose increasingly more complex requirements
on resource and job management software (RJMS).
However, there is a paucity of RJMS techniques
that can solve key technical challenges
associated with those new requirements, particularly 
when they are coupled. In this paper, we propose a novel
dynamic and multi-level resource model
approach to address three key well-known challenges 
individually and in combination: i.e.,
1) RJMS dynamism to facilitate job and workflow
adaptability, 2) integration of specialized external resources
(e.g. user-centric cloud bursting), and
3) scheduling cloud orchestration framework tasks.
The core idea is to combine a dynamic directed graph
resource model with fully hierarchical
scheduling to provide a unified solution to all three key challenges.
Our empirical and analytical evaluations of the solution using our
prototype extension to Fluxion, a production
hierarchical graph-based scheduler, suggest
that our unified solution can significantly
improve flexibility, performance
and scalability across all three problems 
in comparison to limited traditional approaches.
\end{abstract}


\maketitle


\section{Introduction}
\label{sec:intro}

The demise of Dennard scaling in the early 2000s 
spurred high performance computing (HPC) to 
embrace parallelism and rely on Moore's law   
for continued performance improvement.  HPC now 
finds itself near the end of Moore's law. 
The end of the two trends has resulted in 
exponentially increasing parallelism and 
the adoption of accelerators (GPUs, 
field programmable gate arrays, and other 
specialized technologies), leading to ``extreme 
heterogeneity'' and ``dynamic heterogeneity'' 
in system resources~\cite{ascr2018}. 
Furthermore, the increasing integration of cloud computing 
with HPC allows applications on-demand access 
to pools of external resources that can change 
over time depending on their cost 
or availability. State-of-the-art 
supercomputers feature extreme and 
dynamic heterogeneity and incorporate 
cloud technologies, forming highly complex and 
dynamic systems~\cite{sdsc-expanse,stanzione2020,psc-bridges-2}. 
Exascale systems are expected to further  
the trend to a greater scale and integrate cloud 
resources seamlessly. 

Cutting-edge scientific workflows are coevolving and 
becoming far more complex, often integrating machine learning 
and visualization with multiscale or multiphysics models. 
The added sophistication derives from a need to solve large, 
interdisciplinary problems such as automating lead optimization 
in drug discovery~\cite{yang2020}, understanding protein interactions 
with cell membranes for cancer research~\cite{dinatale2019}, 
and responding to global crises like the COVID-19 
pandemic by accelerating drug design~\cite{minnich2020}.
According to the United States Department of Energy,
increasingly complex ``workflows will 
need to adapt on-the-fly to rapidly changing resource,  
data, and user constraints'' and require resource 
management and scheduling adaptability to 
``match the dynamically changing 
resource requirements of a job, as well as the 
changing performance and availability of 
the resources themselves'' and in turn alter 
scheduling decisions in ``response to unfolding 
events such as new data, changes to [a] workflow, 
or changes in cost or availability of resources''~\cite{ascr2018}. 

There are three well-known key technical challenges
one must solve to address extreme and dynamic heterogeneity: 
1) RJMS dynamism to facilitate job and workflow adaptability, 
2) integration of specialized external resources 
(e.g. user-centric cloud bursting), and 
3) scheduling elastic cloud orchestration 
framework tasks.
While each of these challenges is unique and 
individual solutions to each challenge do
exist (e.g.~\cite{prabhakaran2015,lsf,volcano}), we argue that these challenges
must be viewed as ``interrelated": i.e., they are expected
to be increasingly combined as HPC and cloud environments 
become more tightly converged. However, there 
is a dearth of work on techniques 
that consider and address all three core problems combined.

We observe that a lack of more effective resource data models
commonly underlies these problems.
Extreme heterogeneity and resource dynamism 
produce an explosive growth in the space of configurations 
that can be requested by a job. The rapid growth of the 
resource configuration space exposes limitations of existing RJMS.
Current resource data models are often
based on simplistic, rigid representation schemes, which may
also be static, requiring that new resource types and 
relationships be defined in a configuration file.
Static representations are ill-suited 
to satisfying requests for dynamic and external 
(e.g. cloud) resources, and limit 
the ability to integrate resources 
returned from a cloud provider based on 
specification of cost or location rather than hardware. 


We propose novel distributed graph-based
techniques as a dynamic, fully hierarchical resource model 
to solve all three challenges individually and in combination. 
The primitive operations of our solution
are adding or removing a target resource subgraph from 
an existing resource allocation or system.  
Our solution includes a novel technique that ensures
scalability by adding or removing a subgraph without updating 
the entire graph state, which can be distributed 
across a hierarchy of scheduler instances.
Our approach is formulated in the context of 
fully hierarchical scheduling, one of the 
most general forms of HPC scheduling. 
We adapt fully hierarchical scheduling to integrate externally
provided resources (e.g., cloud resources) 
selected and returned by the external provider, 
and enable specialization of the provider itself. 
Our unique strategy of combining fully hierarchical scheduling with a dynamic graph-based resource 
model provides the flexibility and efficiency to manage massive 
resource configuration spaces.

Specifically, we make the following contributions: 

\begin{itemize}
        \item the notion of resource subgraph inclusion partial ordering within fully hierarchical scheduling
              which allows us to define resource graph transformations
              (e.g., adding a new vertex or edge) and 
              external resource specialization;
	    \item embodiment of simple hierarchical and distributed algorithms for resource
              dynamism by extending Fluxion\cite{fluxion}, a production graph-based scheduler; 
	    \item suitability of our solution for local and external resource addition by testing
              its performance on a local cluster, with HPC+AWS and on a converged HPC and cloud environment
              by extending KubeFlux (Kubernetes integrated with Fluxion) with elasticity, 
              and performing an analytical analysis of the hierarchical algorithms.
\end{itemize}

Our studies reveal that there is minimal memory and runtime overhead 
associated with resource addition of local and external cloud 
resource for single-level jobs, and indicate 
that our methods are efficient for multi-level jobs.  


\section{Current Efforts to Manage Dynamic Heterogeneity}
\label{sec:motivation}

Complex workflows such as the Multiscale Machine-Learned 
Modeling Infrastructure (MuMMI)~\cite{dinatale2019},  
Accelerating Therapeutics Opportunities in Medicine  
Modeling Pipeline (AMPL)~\cite{minnich2020}, 
and Clustered Atom Subtypes aidEd Lead Optimization 
(CASTELO)~\cite{yang2020} represent directions 
of future workflow development and 
feature several techniques that depend on resource dynamism 
and heterogeneity. In particular, MuMMI and AMPL 
contain stages with many independent tasks, 
known as ensembles, which are becoming increasingly prevalent in HPC. 
AMPL and CASTELO also rely on external resources 
in the form of on-premises cloud technologies. 
In this Section, we explore current work on resource 
dynamism and limitations and challenges.

\subsection{Ensemble-based workflows and elasticity}

Ensemble jobs constitute as many as 48\% of the jobs run on a large
cluster sited at Lawrence Livermore National Laboratory (LLNL),
one of the world largest supercomputing centers, and 
can severely tax the capabilities of centralized 
schedulers on near-exascale machines~\cite{flux-fgcs:2020}.
Large centers like LLNL use fully hierarchical scheduling to tackle 
the scalability limitations of centralized schedulers. 
In fully hierarchical scheduling, any scheduler instance can spawn child
instances to aid in scheduling, launching, and managing jobs,
which can recurse to an arbitrary depth.
While fully hierarchical scheduling has proven effective in overcoming
scheduling bottlenecks~\cite{flux-fgcs:2020},
researchers at LLNL discovered that
for those ensemble workflows that change resource requirements
at runtime, an inability to resize
resource shapes at lower level schedulers
can still lead to resource underutilization
or limit performance. 

Jobs or workflow components that change shape 
through adding more resources, relinquishing resources, 
or allowing their allocations to be altered by the scheduler can 
increase efficiency and 
throughput~\cite{kale2002,gupta2014,prabhakaran2015}. 
Jobs capable of one or more of these types of dynamism are 
called adaptive jobs \cite{feitelson1996}; 
the Slurm scheduler~\cite{yoo2003} supports adaptive jobs by allowing a
running job to submit a new job with a dependency indicator and then merging the
allocations. However, the approach used by Slurm demands
that an adaptive job release all the nodes
that came with a single dynamic request.
Hereafter we use ``elastic jobs'' to designate adaptive jobs that can 
``burst'' to cloud or other external resources.

\subsection{Converged computing and cloud bursting}
\label{subsec:native-bursting}


Bringing the best features of HPC and cloud 
together in a converged environment is of 
considerable interest to both communities. 
Containers offer significant advantages to traditional 
applications in terms of reproducibility and portability, 
which are attractive to the HPC and data science communities.
Kubernetes is the de facto container orchestrator, and it is widely used 
for running cloud-native applications ranging from traditional 
microservices to AI workflows. Kubernetes is highly customizable, 
and it can be extended to introduce new capabilities. 
HPC RJMS provide sophisticated scheduling based on desired application placement 
and shape such as fine-grained hardware requirements, job co-scheduling,
easy specification of affinity, or group/gang scheduling. Kubernetes' default scheduler does not provide these features, 
so the orchestration community is responding to address these limitations 
by creating new plugins such as NUMA aware container 
placement\footnote{https://kubernetes.io/docs/tasks/administer-cluster/topology-manager/}.
IBM and LLNL's approach the problem through a collaboration to integrate Fluxion into Kubernetes. 
Specifically, the IBM-LLNL KubeFlux project implements a
plugin scheduler for Kubernetes using Fluxion.
To create a job allocation, KubeFlux invokes Fluxion's
\texttt{resource-query} tool with 
a Fluxion job specification that includes an encoded Kubernetes
pod specification. While this addresses the aforementioned HPC scheduling problems
within the cloud, KubeFlux lacks the ability to grow or shrink a resource allocation.
Similarly to typical Kubernetes workflows,
users running scientific workflows under KubeFlux want the capability to grow an allocation
and spawn more containers as needed for some phases or to shrink
the allocation during others. A lack of elasticity within current HPC RJMS 
leads to a technology mismatch that prevents seamless
convergence between HPC and cloud computing.

%

A popular form of HPC and cloud convergence is cloud bursting, 
where an application running on an HPC system extends its execution 
environment into a cloud. Bursting allows applications 
to consume additional resources if no satisfactory resources are 
available in the local HPC cluster. While a hybrid HPC-cloud 
environment can outperform either of its components, 
care must be taken to identify applications which can benefit 
from bursting~\cite{gupta2013}.  Much work has been done studying 
application suitability for cloud execution and bursting, 
which has concluded that network performance is a primary 
bottleneck, impacting tightly-coupled applications in 
particular~\cite{netto2018}. However, more loosely-coupled 
and high-throughput applications (or steps for a composite 
application) are suitable for cloud environments~\cite{netto2018}. 

Several production HPC schedulers such as 
Slurm~\cite{yoo2003}, PBS Pro~\cite{pbspro}, and Spectrum LSF~\cite{lsf} 
support bursting into the cloud with elastically-provisioned resources.
They provide policies defining when to burst into and 
relinquish resources from the cloud and limit the 
maximum number of cloud resources to use at any
time. However, Slurm and PBS Pro possess significant 
limitations to their bursting capabilities. 

Slurm and PBS Pro base their resource data models on 
simplistic, rigid representation schemes such as bitmaps.
A bitmap is a rigid representation of a set of homogeneous compute nodes and 
their states where each bit represents whether a node is allocated or free.
While bitmaps are highly optimized
for rigid HPC resources and workflows (e.g.,
using a few bitwise operators to find idle nodes),
the need for global updates to each dynamic
resource expression and difficulties in handling
diverse resources make this approach increasingly confining. 

Slurm and PBS Pro define resources and their 
attributes via static configuration.
As a result, the cloud instance types 
must be chosen by the user a priori, either by selecting a queue tied to a
particular type or by providing a command line argument
specifying the instance types. 
We hypothesize that such limitations often arise from the rigid
resource models used in the traditional approach. These limitations
not only include difficulties with static configurations but also
performance variabilities when they do not consider
performance-critical constraints for the elastic resources (e.g., the
hosts must be scheduled from a specific number of zones and racks). 

Spectrum LSF is far more flexible in that it can integrate cloud resources 
that are chosen by the cloud provider to meet requests based on location or cost. 
However, it also suffers from limitations derived from static mappings. 
Enabling or reconfiguring users for bursting to cloud providers 
is achieved via a static, cluster-wide configuration. LSF does not  
currently support user-centric external resource specialization.


\section{Scalable Graph Editing Techniques for Elasticity}
\label{sec:approach}

Directed graph resource models and fully hierarchical scheduling 
are powerful tools, but in combination they address the 
significant challenges we identify in Section~\ref{sec:intro} 
of managing extreme, dynamic heterogeneous configuration spaces. 
In this section we define our core approach through a subgraph 
partial ordering and primitives for addition and removal of resources 
within fully hierarchical scheduling. 
Graphical fully hierarchical scheduling is one of the most general ways to address HPC and cloud convergence and 
elastic jobs since the distinction between external and local resources 
is recast as a parent and child relationship. 
In the discussion that follows we assume the scheduling hierarchy 
is a tree. In the context of fully hierarchical scheduling, we refer to the initial 
hierarchical scheduling instance as the top-level scheduler and its resource graph as 
$G_0$.  Note that any child instance of the multi-level scheduler will instantiate 
a resource graph that is a subgraph of $G_0$.  More generally, 
for any child instance resource graph $G_c$ of a parent instance 
resource graph $G_p$: $G_c \subseteq G_p$. 
To reduce memory usage, each instance initializes its resource graph 
with only those resources within its purview.  As a simple example, 
for a cluster of three nodes, an instance that has requested 
and received two nodes from the top level will instantiate 
its resource graph with only those two nodes. The child has 
no knowledge of resources outside of its graph.  For a general 
nested job of $n$ levels with each level starting a different 
number of child jobs $j = (a_1, a_2, \ldots a_k)$, we have a 
sequence of sequences of jobs $(j_i)_{i=1}^n$. If each child 
starts exactly one job, we have the following sequence of 
resource subgraph partial orderings: 

\begin{equation}
G_0 \supseteq G_{1} \supseteq G_{\sm{1}_{\vsm{1}}} 
\supseteq \ldots \supseteq G_{\underbrace{\sm{1}_{\vsm{1}_{\ddots_1}}}_{n\text{-times}}} 
\end{equation}

Now a dynamic resource subgraph is any resource subgraph that can 
change its topology.  A typical example is to add a new vertex and 
edge.  There are several implications to consider.

First, subjecting any subgraph in the sequence to a 
nontrivial topology transformation invalidates the subgraph inclusion sequence. 
In particular, if some graph transformation $T$ is applied to 
any subgraph in the inclusion sequence,
the entire sequence can be invalidated. 
More formally, a general $T$ which adds new and deletes existing 
vertices and edges invalidates the entire inclusion sequence:

\begin{equation}
G_0 \not\supseteq G_{1} \not\supseteq T(G_{\sm{1}_{\vsm{1}}}) 
\not\supseteq \ldots
\end{equation}

Second, a transformation that only adds vertices and edges invalidates the 
supergraph inclusion subsequence, while a transformation 
that only deletes vertices and edges invalidates the subgraph 
inclusion subsequence. The invalidation ``direction'' imposes a 
natural direction of application for the transformation operation $T$. 
Therefore an additive transformation naturally proceeds from the 
top down, while a subtractive transformation moves from the bottom 
up. Third, in order to add vertices to subgraph 
$n$, the vertices must be added to the $n-1$ 
supergraphs. The addition is the identity if the vertices 
already exist in any supergraph.  A similar argument can be 
applied to the removal of vertices or edges.
Fourth, which vertices and edges must be added
are predicated on finding resources {\em matching} a resource request.

External resource specialization, or the ability 
for each user to choose the external resource 
providers of their preference is related to the 
subgraph partial order of graphical fully hierarchical scheduling. For 
external resource specialization 
we allow the additive transformation 
to invalidate the supergraph inclusion sequence. 
In other words, external resources $E_i$ 
are managed by a first-level allocation 
$G_i$ independent of the top-level scheduler. 
Independent management of external resources 
cannot create allocation inconsistencies since 
$G_i \backslash G_0 = E_i$. Independent 
management enables consistent resource  
management for bursting to external resources as well 
as full flexibility for specializing the external 
provider. However, if a site requires that external 
resources are available for any user the 
addition transformation can be configured not to invalidate 
the supergraph inclusion sequence.
Graph-based fully hierarchical scheduling is naturally suited for pairwise 
parent-child communications between levels. The modular 
parent-child structure enables a simple recursive 
strategy for communicating subgraph changes. 
Resource transformations are communicated up or down 
the subgraph sequence, and are prompted by a 
resource match request specification 
referred to as a jobspec. The procedure 
to match available resources to a jobspec 
and allocate them upon a successful 
match is \textproc{MatchAllocate(MA)}. \textproc{MatchGrow(MG)} 
attempts to match available resources 
to a jobspec in the local subgraph, 
but forwards the request to its parent 
if the match fails. See Algorithm~\ref{algo:butd-grow-add} 
for details.

In the following discussion we assume that a leaf 
scheduler instance makes a request for additional 
resources, but any nested level can initiate a 
request. To add resources to a leaf instance, the leaf 
issues a \textproc{MG}, passing a jobspec 
as an argument. If satisfactory resources are located 
within the leaf's resource graph, the resource graph 
scheduling metadata is updated to indicate that new resources are 
now attached to the job. A successful single-level 
\textproc{MG} behaves almost identically 
to the standard \textproc{MA}; the
difference is that the new resources are 
given the allocation metadata of a running 
job allocation. If resources are not found locally, 
the leaf issues a \textproc{MG} to its parent.
If the new resource request cannot be met, 
the parent issues the same job request to its parent  
and so on up the tree via \textproc{MG} recursion.
When matching resources are found, the scheduling 
metadata is updated and the new subgraph is sent to the child. 
The child adds the new vertices and edges into its resource 
graph and updates its scheduling metadata. The 
top-down process proceeds until reaching the 
originating leaf. We refrain from discussing 
procedures for removing resources from the 
subgraph inclusion sequence since they
are analogous to \textproc{MG}.

To be scalable, it is important to limit the requisite
updates to the minimum set of vertices and edges
in these procedures. Thus, we use a technique called
{\em localization} as we only perform local
updates in these procedures. 
Assuming the graph implementation indexes vertices by their 
graph paths, which holds true for the Fluxion implementation of 
fully-hierarchical scheduling, 
we can use the index to locate the attaching point of the 
subgraph in constant time. This makes the computational complexity
of \textproc{AddSubgraph} 
$\mathcal{O}(n + m)$ for a subgraph
of $n$ vertices and $m$ edges.
Further, the metadata within each vertex is organized
such that each vertex will only contain
the metadata about itself and certain quantities
as a function of its subgraph (e.g., subtree
rooted at that vertex). Thus,
attaching a new subgraph only requires 
updating its ancestor vertices.
This limits the computational complexity
of \textproc{UpdateMetadata} 
to be $\mathcal{O}(n + m + p)$ for
a subgraph of $n$ vertices and $m$ edges with $p$ supergraph ancestors.

We provide an \textproc{ExternalAPI} 
to translate additive or subtractive transformations 
from the hierarchical scheduler into external resource 
provider functions. The \textproc{ExternalAPI} 
takes a jobspec as an argument, and 
translates the specification to a request for  
external resources, calling the provider API 
to create or start instances.  The API 
returns the resources as a subgraph, so the remainder 
of the \textproc{MG} process can proceed as if the 
resources originate from a local parent. 
To a scheduler instance, the external 
resource provider is functionally just 
another parent in the hierarchical scheduling.

\begin{algorithm*}[htb]
    \begin{algorithmic}[1]
    \Function{AddSubgraph}{$subGraph$}
        \State $graph \leftarrow \textproc{GetGraph}()$
        \For {$(sourceVertex, targetVertex)$ in $subGraph.edges$}
            \If {$sourceVertex$ and $targetVertex$ in $graph.vertices$}
                \If {$(sourceVertex, targetVertex)$ not in $graph.edges$}
                    \State $\textproc{AddEdge}(graph, sourceVertex, targetVertex)$ \Comment{graph library native function}
                \EndIf
            \Else 
                \If {$sourceVertex$ not in $graph.vertices$}
                    \State $\textproc{AddVertex}(graph, sourceVertex)$ \Comment{graph library native function}
                \EndIf
                \If {$targetVertex$ not in $graph.vertices$}
                    \State $\textproc{AddVertex}(graph, targetVertex)$
                \EndIf
                \State $\textproc{AddEdge}(graph, sourceVertex, targetVertex)$
            \EndIf
        \EndFor
        \Return
    \EndFunction

    \Function{RunGrow}{$subGraph, add$}
        \If {$add$} \Comment{add $subGraph$ into $graph$}
            \State $\textproc{AddSubgraph}(subGraph)$
        \EndIf
        \State $\textproc{UpdateMetadata}(subGraph)$ \Comment{Existing Fluxion function to update scheduler state}
        \State \Return
    \EndFunction

    \Function{MatchGrow}{$jobSpec$}
        \State $subGraph \leftarrow \textproc{MatchAllocate}(jobSpec)$
        \If{$subGraph$} \Comment{resources found locally}
            \State $\textproc{RunGrow}(subGraph, false)$
        \Else \Comment{$subGraph = \varnothing$}
            \State $parent \leftarrow \textproc{GetParent}()$
            \If{$parent = \varnothing$} \Comment{top level}
                \If {$\nexists \textproc{ExternalAPI}$}
                    \State \Return $\varnothing$              
                \EndIf
                \State $subGraph \leftarrow \textproc{ExternalAPI}(jobSpec)$
            \Else 
                \State $subGraph \leftarrow \textproc{RemoteProcedureCall}(parent, \textproc{MatchGrow}(jobSpec))$
            \normalsize\EndIf
            \State $\textproc{RunGrow}(subGraph, true)$
        \EndIf
        \Return subGraph
    \EndFunction
    \end{algorithmic}
    \caption{Bottom-up then top-down MatchGrow procedure}
    \label{algo:butd-grow-add}
\end{algorithm*}


\section{Enabling Elasticity in Fluxion}
\label{sec:implementation}

In this section we describe our implementation of
the procedures described in Section~\ref{sec:approach} 
in Fluxion to enable elasticity and 
management of extreme, dynamic heterogeneity.
We also discuss how enabling elasticity 
in Fluxion allows cloud computing technology 
like KubeFlux to grow or shrink resource allocations.

The Fluxion resource graph relies on Boost 
Graph Library (BGL)~\cite{bgl2002} to 
provide the underlying directed graph data 
structure. We use vertex and edge addition 
and removal functions provided by BGL to 
transform the resource graph. Subgraphs 
to be added or removed are encoded in 
JSON Graph Format (JGF) which can then 
be transmitted between parent and child 
schedulers via Remote Procedure Call (RPC)
functionality built into the Flux RJMS 
framework~\cite{flux-fgcs:2020}. For example, if a 
child issues \textproc{MatchGrow} 
via RPC to its parent and the match is successful, 
the matching resources are returned in JGF as part 
of the call.  

To add and remove resources from an external 
provider, we implement \textproc{ExternalAPI} 
from Section~\ref{sec:approach} with the popular AWS API.
The AWS API is extremely rich and configurable, 
making it an ideal API for use with Fluxion.
We use the Python ({\tt boto3}) AWS API as it is 
well-documented and suitable for 
rapid integration prototyping in Fluxion. Our implementation 
(\textproc{EC2API}) takes a Fluxion jobspec as an input argument, 
and depending on the jobspec either maps the request 
to corresponding EC2 instance types or builds an EC2 
Fleet request for generic resources. Then 
\textproc{EC2API} makes the AWS API call to 
create the resources and returns the new 
subgraph in JGF to Fluxion.  
One of the powerful capabilities afforded by a directed 
graph representation of resources is the inclusion of the physical 
location of the resources.  \texttt{EC2API} can 
interpose an EC2 zone vertex between the nodes' vertices 
and the cluster vertex. Location information allows the 
application to make location-dependent decisions which 
can be particularly important for using cheap AWS 
spot instances or reliably storing data in multiple S3 zones.


\section{Experimental work}
\label{sec:experiments}

In this section we conduct a series of experiments 
to test whether Fluxion, a graph-based fully hierarchical scheduler, can effectively 
provide the three capabilities identified in 
Section~\ref{sec:intro} as necessary for managing 
extreme, dynamic heterogeneity. 
First, we investigate the performance of additive graph 
transformation in the context of a traditional 
adaptive HPC task. Second, we test Fluxion's performance 
for scheduling elastic Kubernetes tasks with KubeFlux. 
Finally, we examine Fluxion's \textproc{ExternalAPI} performance 
by integrating AWS cloud resources, and compare its 
flexibility to that of Slurm and Spectrum LSF through 
managing EC2 Fleet resources. The experiments in this section test the three
primary components of resource graph growth: the time to
find new resources, the time to communicate new
resources in the form of a subgraph, and the
time needed to add the new subgraph into the
existing resource graph and update its state. 

We test Fluxion on two nodes of a dedicated
test cluster at LLNL. Each node features dual-socket Intel Xeon E5-2660 CPUs
with 16 cores and 64 GB memory; internode communications
are performed across Mellanox QDR HCAs via
IP over InfiniBand (IPoIB). We populate test graphs with the configurations
detailed in Table~\ref{tab:level-specs} with a tree topology, and the test jobspecs
are described in Table~\ref{tab:test-specs}.
We use the default Fluxion settings except for the pruning filter setting, for which we
use \texttt{ALL:core}, because it ensures better graph search pruning for our tests.
We test KubeFlux on a 26 node OpenShift cluster at IBM T.J. Watson research center. 
Each node contains 2 sockets with 10 Power8\texttrademark~per socket and SMT8 for a
total 160 cores per node, 4 Tesla K80 GPUs, and 512 GB memory.

\subsection{Single-level overhead}
\label{subsec:sl-ov}

We first evaluate the performance and memory
overheads at the single-scheduler level.
We employ \texttt{resource-query},
a Fluxion utility for scheduler testing and reporting
resource graph and job allocation statistics.
We run our tests on a single node, starting
one \texttt{resource-query} process at a time.

To establish a comparison between \textproc{MatchGrow} (\textproc{MG})
and \textproc{MatchAllocate} (\textproc{MA}) (Sec.~\ref{sec:approach}), 
we test two different resource graphs.
The baseline test configuration
begins by initializing a resource graph of 143 vertices
and edges (L3 in Table~\ref{tab:level-specs}).  
The test issues two \textproc{MA}
calls with each jobspec corresponding to T7 in
Table~\ref{tab:test-specs}. The \textproc{MG} test
configuration initializes a subgraph of 73 vertices 
(L4 in Table~\ref{tab:level-specs})
and edges and issues a \textproc{MA} requesting
all resources (one node with two sockets and 32 cores).  
Once satisfied, the test executes an 
\textproc{MG} requesting a subgraph corresponding
to T7.  When the subgraph is added into the local resource
graph, the second test's resource graph contains the same
vertices and edges as the first, but has a single allocated job.
For each test, \texttt{resource-query} reports the
max Resident Set Size (RSS) after initialization and before test finalization.
We also record the elapsed time of \textproc{MA} 
and \textproc{MG}.  We repeat each test
100 times and compute the average timings.
The results indicate that the average match times for both 
tests are approximately equivalent (0.002871s for 
\textproc{MA} versus 0.002883s for \textproc{MG}),
but the subgraph update procedure takes 0.005592s on average for \textproc{MG} (\textproc{MA} 
does not require the update).
The max RSS values are also
comparable (5776kB for \textproc{MA} versus 5840kB for \textproc{MG}), indicating that \textproc{MG} increases
memory usage linearly in the subgraph size.

\subsection{Nested \textsc{MatchGrow}}
\label{subsec:nested-mg}

We run nested \textproc{MG} tests on two nodes with
a hierarchy of five levels.  Each level in the
hierarchy populates a resource graph in
JGF with identical Fluxion scheduling
options.  Concretely, L0 is run on
physical node0, and reads a JGF to populate a
simulated cluster-level resource graph consisting 
of 18,061 vertices and edges.  Level1
along with levels two through four reside on node1,
which is configured identically to node0. 
The deeper nested levels are configured similarly.
Table~\ref{tab:level-specs} shows
each level's constituents and graph size.
The experiments test the timings for each level to issue an
RPC and to receive a subgraph in return, perform an 
\textproc{MG}, add the received subgraph
into the local resource graph, and update the scheduler
metadata. We perform six tests with successively larger
requested subgraph sizes.  Subgraph sizes are listed by
their number of vertices plus edges. For example, the
final test (T8) is a jobspec request of 16 cores on one
socket (equivalent to the requests previously allocated
at this level during test initialization). See 
Table~\ref{tab:test-specs} for request details.

\begin{table}[htbp]
\caption{\textsc{MatchGrow} request tests}
\centering
\scriptsize
\begin{tabular}{|c|c|c|c|c|}
\hline
\textbf{Test} & \textbf{nodes} & \textbf{sockets} & \textbf{cores} & \textbf{graph size}\\
\hline
\textbf{T1} & 64 & 128 & 2048 & 4480 \\
\hline
\textbf{T2} & 32 & 64 & 1024 & 2240 \\
\hline
\textbf{T3} & 16 & 32 & 512 & 1120 \\
\hline
\textbf{T4} & 8 & 16 & 256 & 560 \\
\hline
\textbf{T5} & 4 & 8 & 128 & 280 \\
\hline
\textbf{T6} & 2 & 4 & 64 & 140 \\
\hline
\textbf{T7} & 1 & 2 & 32 & 70 \\
\hline
\textbf{T8} & 0 & 1 & 16 & 36 \\
\hline
\end{tabular}
\label{tab:test-specs}
\end{table}

\begin{table}[htbp]
\caption{Graph sizes by level}
\centering
\scriptsize
\begin{tabular}{|c|c|c|c|c|}
\hline
\textbf{Level} & \textbf{nodes} & \textbf{sockets} & \textbf{cores} & \textbf{graph size}\\
\hline
\textbf{L0} & 128 & 256 & 4096 & 18,061 \\
\hline
\textbf{L1} & 8 & 16 & 256 & 563 \\
\hline
\textbf{L2} & 4 & 8 & 128 & 283 \\
\hline
\textbf{L3} & 2 & 4 & 64 & 143 \\
\hline
\textbf{L4} & 1 & 2 & 32 & 73 \\
\hline
\end{tabular}
\label{tab:level-specs}
\end{table}

After the environment is completely initialized, a helper script starts
a test by executing an \textproc{MG} at the leaf (L4), specifying
one of the test requests in jobspec format. Levels1-4 are configured to be fully allocated,
so they successively forward the request up the tree until reaching L0.  L0
succeeds in matching the resources, and returns them to L1 after updating
its scheduler metadata with the new resources allocated to the job.  Upon
completion of the subgraph add and update steps, the helper script reinitializes
the resource graphs at each level and repeats the request.  We perform each test
100 times to gather representative data and compute statistics.

\subsubsection{Inter-level communication time}
\label{subsubsec:interlevel}

A graph-based fully hierarchical scheduler needs to transmit 
subgraphs between child-parent pairs in linear time with no other dependencies.
If a child issues an \textproc{MG} RPC with a particular jobspec and
receives the matched resources from its parent, the 
communication time should only depend on the number of vertices 
and edges in the matched subgraph.

\begin{figure*}[t]
    \begin{subfigure}[t]{0.4\textwidth}
    \includegraphics[width=\linewidth]{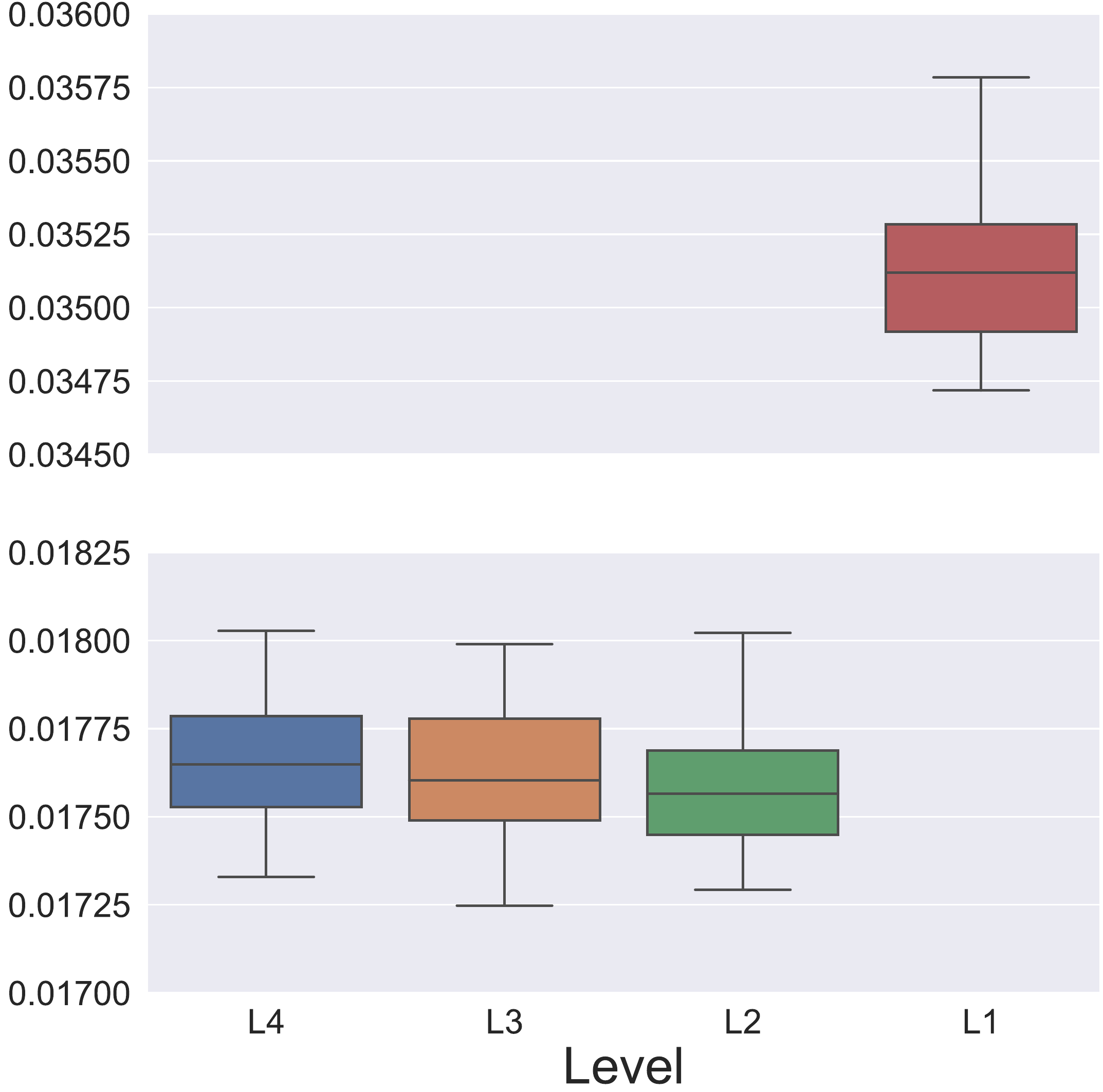}
    \caption{Measured communication timing distributions for test T2, or subgraph
    size 2240.  Each boxplot consists of 100 communication timing measurements.
    Levels2-4 manifest very similarly, with overlapping IQRs and nearly
    equivalent medians.  However, level1 is distinct: its spread is larger, and the
    median value is larger.  Its behavior is different from the deeper nested levels.}
    \label{fig:comms-2240}
    \end{subfigure}
    \hspace{2mm}
    \begin{subfigure}[t]{0.4\textwidth}
    \includegraphics[width=\linewidth]{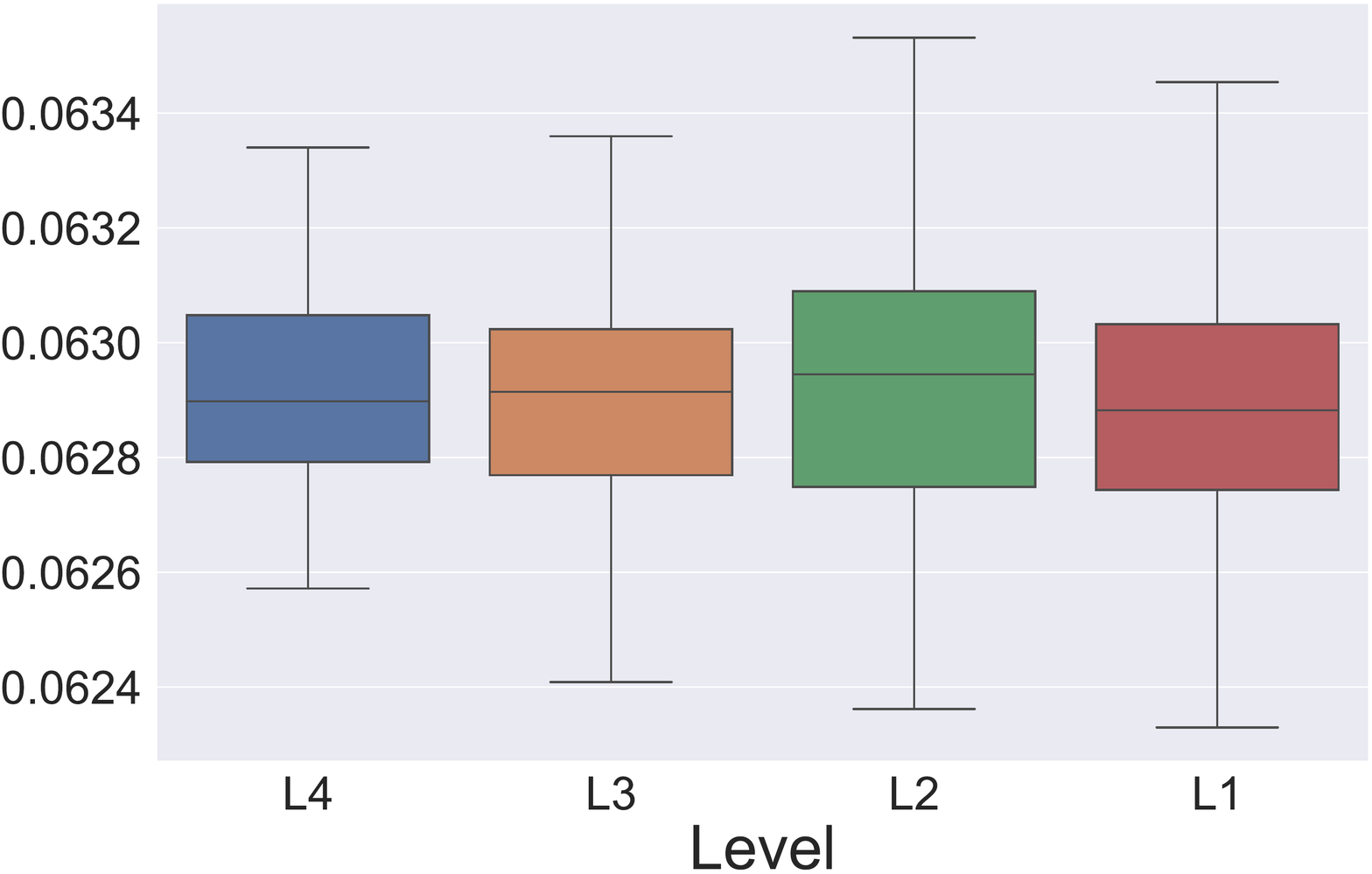}
    \caption{Measured add and update timing distributions for T2, or subgraph
    size 2240.  Each boxplot consists of 100 add and update timing measurements.
    All levels manifest very similarly, with overlapping IQRs and nearly
    equivalent medians.}
    \label{fig:attach-2240}
    \end{subfigure}
    \caption{\textproc{MatchGrow} overheads}
\end{figure*}

To measure communication times between child and parent, we record elapsed time between
issuing an RPC and receiving the response.
Figure~\ref{fig:comms-2240} shows the distributions of communication 
times for the T2 test.  In the interest of
clarity, we limit our presentation to this size, since the other
tests' distributions manifest similarly except for their positions
on the vertical axis.  In other words, while the tests indicate that
communication depends on the subgraph size, the behavior across
the tests sizes is similar.

Figure~\ref{fig:comms-2240} demonstrates that
communication time does not depend on resource graph size (implicit
in the levels). Levels two through four are very similar: their spread, as seen
by the interquartile ranges (IQRs) and whiskers, and medians are nearly identical.
However, noting the break in the vertical axis, communication time and behavior
does depend on relative location of the parent and child pair.
The L1 communication time has a markedly wider
IQR, and a much higher median value.  Since L1's parent
resides on a remote node, the communication times include internode
communication, which we expect to be slower than intranode (as
seen with levels two through four).  We build linear 
regression models for inter- and intranode 
communications and evaluate them based on timing data, 
and conclude that the times depend on the request size.

\subsubsection{Subgraph update time}
\label{subsubsec:add-update}
For a resource graph with a tree topology, add and update stages
necessitate an addition operation for each subgraph vertex
and edge, and then a depth-first traversal of the new subgraph 
to update metadata (see Section~\ref{sec:approach}).
In Fluxion, the traversal can be pruned because all previous
resource vertices are allocated, so only the newly added subgraph
needs to be explored.  The add and update stages should only depend on the 
subgraph size and the number of ancestors of the subgraph root.

We measure the time taken to perform the
add and update stages within the overall \textproc{MG} operation.
Figure~\ref{fig:attach-2240} shows the distributions of results for
the T2 test.  Like the communication time tests, the
add-update tests exhibit similar behavior across test sizes;
in the interest of clarity we only present one size here. The
boxplots indicate no dependence on the level's
resource graph size since there are no clear differences between
the IQRs and medians relative positions (e.g. non-overlapping). 

\subsubsection{Resource match time}
\label{subsubsec:match}

The final measurement collected during each test is the time for
each level to attempt to find resources matching the request.  For
an abstract graph-based RJMS, we expect the complexity to be
$\mathcal{O}(n + m)$ for a resource graph of $n + m$ vertices and
edges. However, as Fluxion employs pruning filters,
which prevent traversal into allocated subtrees,
the complexity is less. In particular, we expect to see complexity
dependent on the number of high-level resources (higher than core,
at least) in the resource graph.
The results of mean match times across all test
sizes for levels one through four indicate a small dependence
on resource graph size within a test but, there is
no obvious dependence on request subgraph size.
Level one exhibits different behavior as it finds 
successful matches. For a successful match, there is a 
complicated dependence on the request subgraph size and 
topology which is due to the relationship between the pruning filters used and 
the configured level of detail in the resource graph \cite{fluxion}.
Exploring the performance of successful matches 
is not our goal in this work.


\subsection{Bursting to EC2}

\begin{figure}[!thb]
    \centerline{\includegraphics[width=0.45\textwidth]{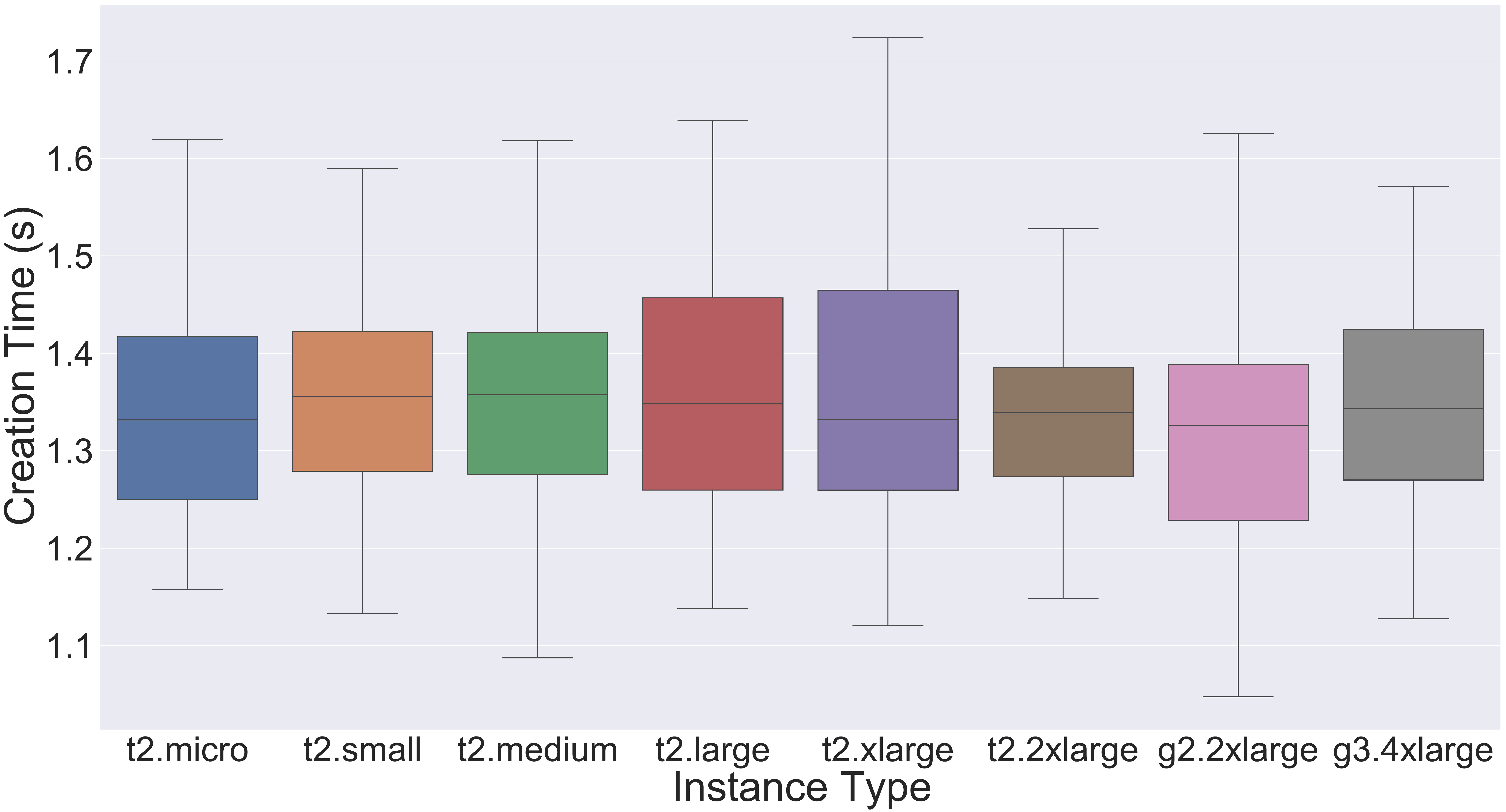}}
    \caption{Boxplots of EC2 creation times}
    {The vertical axis designates the time taken for AWS to satisfy a request test
    by returning a list of instance objects corresponding to the API request.
    Each box plot aggregates the results from 1, 2, 4, and 8 simultaneous
    instance requests of each type.  Therefore, each boxplot contains 80 test results.}
    \label{fig:ec2-box}
\end{figure}

In this section, we test whether a graph-based
RJMS can effectively request and create cloud resources within a reasonable amount
of time.  We also compare the static resource configuration used 
by a traditional bitmap-based scheduler to the Fluxion dynamic resource model
by bursting tests to Amazon EC2 Fleet. We choose AWS EC2 and EC2 Fleet 
for external resource testing due to their rich APIs and tremendous request 
flexibility. 
In our first test we measure the time necessary to request specific instances from EC2 and receive a
response, and the time taken by the Fluxion \textproc{EC2API} plugin to encode the response objects
into a JGF subgraph.  To do so, a script issues \textproc{MG} from
within Fluxion running on a node of our dedicated cluster.
We configure \textproc{EC2API} to support eight different instance types and 
measure the time to request and transform one, two, four, and eight instances
at once. We repeat the test 20 times per instance type and request
size, totaling 640 tests. The time needed for EC2 to satisfy instance creation requests is
effectively constant for all instance types and request sizes up to eight
simultaneous instances. See Figure~\ref{fig:ec2-box} for measured EC2 instance 
creation time aggregated by instance type.
The time to map a Fluxion jobspec to an
EC2 instance creation request, as it requires much less than 1\% of the
instance creation time. The EC2 plugin requires an average overhead to generate JGFs 
from instance objects of 1.6\% of the time needed to create instances. 

To demonstrate the flexibility of dynamic resource binding, we compare Fluxion and
the \textproc{EC2API} with Slurm's static binding through EC2 Fleet requests. EC2 Fleet enables 
requests for sets of instance types, including 
On-Demand and Spot instances. AWS processes the user request specification 
and returns a set of instances that meet the constraints. 
In general, the user does not know which instance types will 
meet the request or their locations, which is readily accommodated 
by dynamic binding. Fluxion is capable of adding any combination 
of instances and their locations into its directed graph 
at runtime, so Fleets are requested via appropriate 
AWS API calls in \textproc{EC2API}.  As a simple test, we 
made 10 Fleet requests of 10 instances each, which took 
an average of 6.24 seconds from request to the 
successful addition of the subgraph (44 vertices and edges each) into Fluxion.  
Note that we allowed AWS to return any of 300 instance types, 
since the AWS API returns an error if all 349 are specified.

Slurm does not support EC2 Fleet, and its reliance on static 
binding results in an unmanageable configuration. 
Encoding EC2 instance types (300 for comparison with Fluxion) and locations (77 current 
Availability Zones) yields 23,100 node types in the Slurm 
configuration.  Furthermore, bursting to multiple instances of the 
same type requires a range of instances per type in a partition. 
Configuring 128 instances per node type as given in the 
Cloud Scheduling Guide~\cite{slurm-csg} results in a partition 
with 2,958,600 nodes. To simulate how Slurm handles such 
a configuration file, we start \texttt{slurmctld} and 
\texttt{slurmd} on an LLNL HPC compute node with the file 
specified. \texttt{Slurmctld} and \texttt{slurmd} consume 100\% CPU each 
and time out for user input (e.g. \texttt{sinfo}) until 
the daemons are killed by the debug queue one hour walltime.

IBM Spectrum LSF Resource Connector is a highly adaptable 
interface for LSF, and supports specific Amazon EC2 instance 
requests as well as bursting to Fleet. Like Fluxion and \textproc{EC2API}, the 
Resource Connector can integrate any EC2 instance 
types returned by Spot Fleet, but unlike Fluxion it does not support 
On-Demand Fleets. Fluxion can make scheduling decisions based 
on Availability Zones returned by Fleet. However, based on Resource Connector documentation 
it is unlikely LSF can further enforce global constraints
on dynamically provisioned resources
(e.g., scheduling hosts from a specific number of zones and racks) unless 
it uses a graph resource representation similar to our proposed solution.
We cannot make further determinations because this is vendor proprietary information.
Furthermore, the Resource Connector exhibits static mapping of cluster users to AWS 
users, so allowing new users to burst to EC2 requires 
reconfiguring the user mappings in the connector. Fluxion 
can avoid this problem by using fully hierarchical scheduling, where 
a nested Fluxion scheduler can use \textproc{EC2API} as a 
specific AWS user. 
See Table~\ref{tab:ec2-test} for details on the
instance configurations and equivalent subgraph sizes.

\begin{table*}[tbp]
    \caption{EC2 request tests}
    \begin{center}
    \begin{tabular}{|c|c|c|c|c|}
    \hline
    \textbf{instance type} & \textbf{CPUs} & \textbf{memory (GB)} & \textbf{GPUs} & \textbf{subgraph size}\\
    \hline
    \textbf{t2.micro} & 1 & 1 & 0 & 6 \\
    \hline
    \textbf{t2.small} & 1 & 2 & 0 & 8 \\
    \hline
    \textbf{t2.medium} & 2 & 4 & 0 & 14 \\
    \hline
    \textbf{t2.large} & 2 & 8 & 0 & 22 \\
    \hline
    \textbf{t2.xlarge} & 4 & 16 & 0 & 42 \\
    \hline
    \textbf{t2.2xlarge} & 8 & 32 & 0 & 82 \\
    \hline
    \textbf{g2.2xlarge} & 8 & 15 & 1 & 42 \\
    \hline
    \textbf{g3.4xlarge} & 16 & 128 & 4 & 282 \\
    \hline
    \end{tabular}
    \label{tab:ec2-test}
    \end{center}
\end{table*}
    
\vspace*{-4mm}

\subsection{KubeFlux}

Finally, we tested KubeFlux (see Section~\ref{subsec:native-bursting}) 
performance on an OpenShift cluster as described above. 
KubeFlux is composed of three main parts: 1) Fluxion management level, 2)
Fluxion daemons (FluxRQ), and 3) the resource graph. The
management level collect information to build the
resource graph, listens to new scheduling requests, 
creates job requests to submit to FluxRQ instances, and
defines how the resource graph is partitioned among FluxRQ instances. 
With a new scheduling request, the management level sends a message 
containing jobspec to a FluxRQ instance. FluxRQs pods run gRPC servers, which wait for pod binding requests on
the partition of the Kubernetes cluster described in their resource graph. Upon receiving a 
binding request, FluxRQs build the Fluxion jobspec in YAML format and
submits a \textproc{MA} allocation query to get the target node for pod binding.
As part of this effort, we extended KubeFlux so that
it can make requests to add and remove resources using \textproc{MG}.
The test resource graph contains 4344 vertices and 8686 edges. We evaluate \textproc{MA}
and \textproc{MG} when deploying a Kubernetes ReplicaSet with a single pod first,
and then scale it up to 100 pods. The \textproc{MA} request is executed
once for the first pod allocation, and the average execution time is
0.101810s, while the average execution time for \textproc{MG} requests is
0.100299s.


\section{Analysis and modeling}
\label{sec:analysis}

In Section \ref{sec:experiments} we asserted that the time 
to complete a child-initiated \textproc{MatchGrow} can be 
described by the combination of three components: the time 
to locate resources, the time to communicate the resource 
request to the parent Fluxion instance and receive a new 
subgraph in response, and the time to add the new subgraph 
into the local instance's resource graph and update its 
job metadata. As each of the components is independent 
of the others, the overall behavior of \textproc{MatchGrow} 
is described simply by the sum of the independent 
components over $n$ levels, i.e. 
$t_{MG} = \sum_{i=0}^{n-1} t_{match_i} + t_{comms_i} + t_{add\_upd_i}$. 

Timing data collected in the experiments 
described in Section~\ref{sec:experiments} supports 
our assertion. The sum of the three 
\textproc{MatchGrow} components at levels one through 
four accounts for 99.8\% of the total measured elapsed
time for these levels. 
Level zero is a special case as 
it does not include the $t_{comms}$ component, 
and it emits the matched resources in JGF.  Since we 
do not measure the time to emit the matching 
subgraph, on average $t_{match_0}$ accounts for 
96.4\% of the total level time.  Across all levels, 
$t_{match} + t_{comms} + t_{add\_upd}$ accounts 
for 98.2\% of the measured time, so we 
proceed with the confidence that by modeling 
each component we account for the 
vast majority of total observed 
\textproc{MatchGrow} runtime. With 
sufficiently accurate component models,
we have a sufficiently accurate 
aggregate model of \textproc{MatchGrow} behavior.

\begin{figure}[!thb]
\centerline{\includegraphics[width=0.45\textwidth]{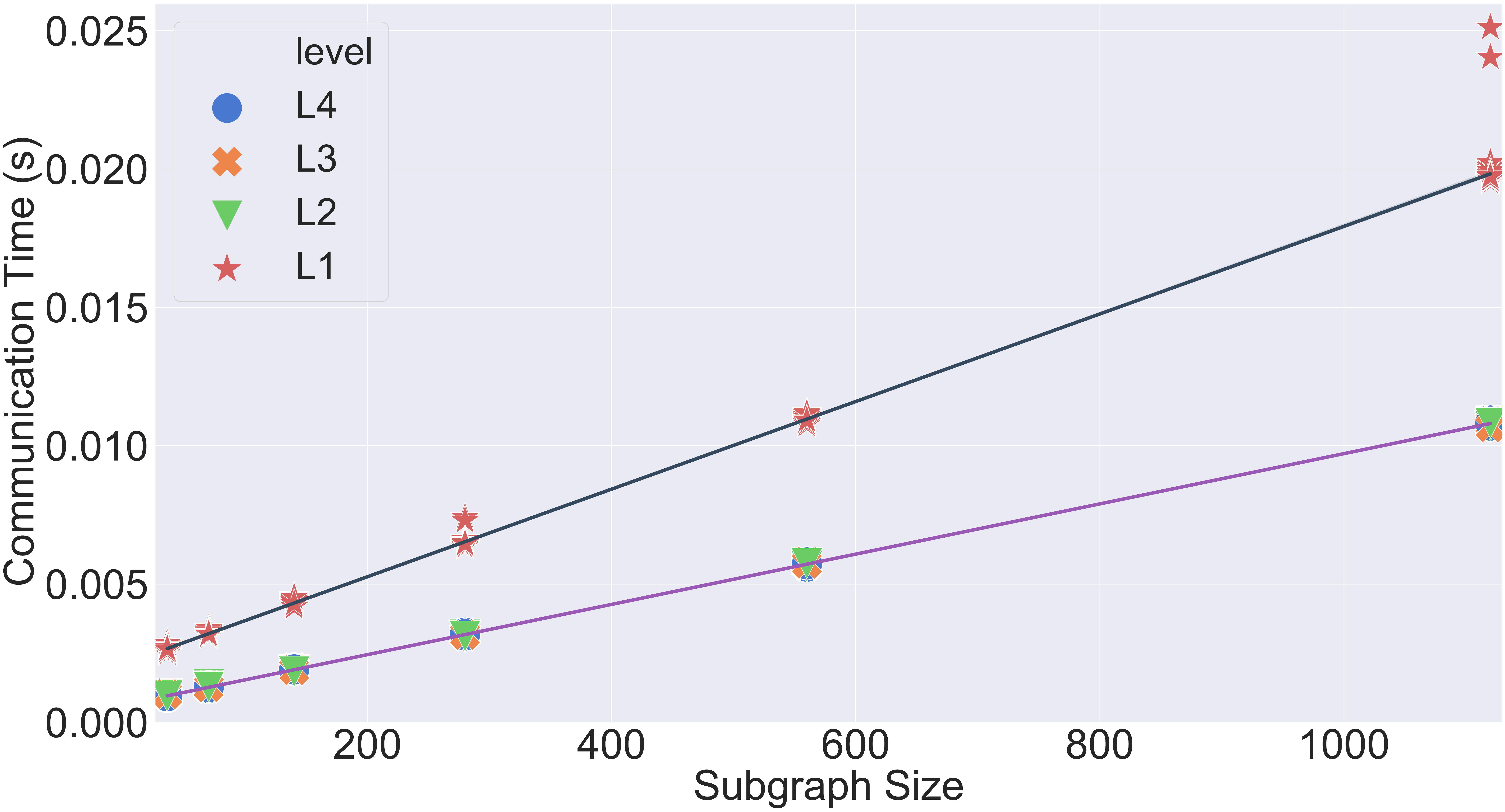}}
\caption{Inter-level communications times by
subgraph request size.  The horizontal axis indicates the size of the
subgraph requested, and the vertical axis indicates the time to communicate
the request to each level's parent via RPC and receive the response as a JGF.
Note levels2-4 obey a different underlying model than level1; the behavior
is due to the different inter-level communications.  The regression models
are plotted as well to give an indication of the models' performance.}
\label{fig:comms-reg}
\end{figure}

\subsection{Communications model}
\label{subsec:comms-model}

Recall from Section~\ref{subsubsec:interlevel} that the inter-level 
communication time should have a linear dependence on the number 
of vertices plus the number of edges ($n$) in the subgraph 
transmitted between levels.  Figure~\ref{fig:comms-2240} indicates 
that there are two such linear models: one for internode subgraph 
communication, and one for intranode subgraph communication:
$t_{inter} = n\beta_{inter} + \beta_{0inter}$, and 
$t_{intra} = n\beta_{intra} + \beta_{0intra}$. We have two regression 
models that we validate with a typical five-fold cross-validation. 
To develop the models and perform the cross validation, 
we use the popular Python package Scikit-learn \cite{scikit-learn2011}.
We report Mean Absolute Percentage Error (MAPE) values, which is a standard 
choice of interpretable error, and the $R^2$ value. 
Table~\ref{tab:reg} lists the MAPE and $R^2$ values averaged across the five 
cross validation tests, and the regression coefficients obtained 
by fitting all data.  Both the inter and intranode models fit the 
data very well, with small average MAPEs and high $R^2$ values. 
Taken together the metrics provide strong evidence that the 
linear models represent the data without omitting independent variables.

Figure~\ref{fig:comms-reg} is a plot of the two regression models 
with data included, and is complementary to Figure~\ref{fig:comms-2240}.  
The vertical axis indicates the time to communicate the request
to each level’s parent via RPC and receive the response as a JGF. 
The horizontal axis indicates the size of the subgraph requested.
The plot provides visible confirmation 
for the excellent model fits, and indicates the different 
$\beta_0$ coefficients returned by regression.  The $\beta_0$ coefficients 
represent the time needed to transmit a subgraph of zero vertices 
and edges, which can be considered the communication initialization 
time.  The $\beta$ coefficients, or slopes, represent the bandwidth 
of each model.  If we tested transmission of much larger subgraphs (millions 
of vertices and edges), we would expect the two models to diverge 
in terms of bandwidth: the intranode bandwidth should exceed 
the internode bandwidth as IPoIB becomes a bottleneck. 
To apply these models, we must know whether a parent and child communicate 
locally or remotely. However, we assume we 
are free to choose the Fluxion job hierarchy which holds 
whenever a user controls its placement.

\subsection{Add-update model}
\label{subsec:add-upd-model}

\begin{figure}[!thb]
\centerline{\includegraphics[width=0.45\textwidth]{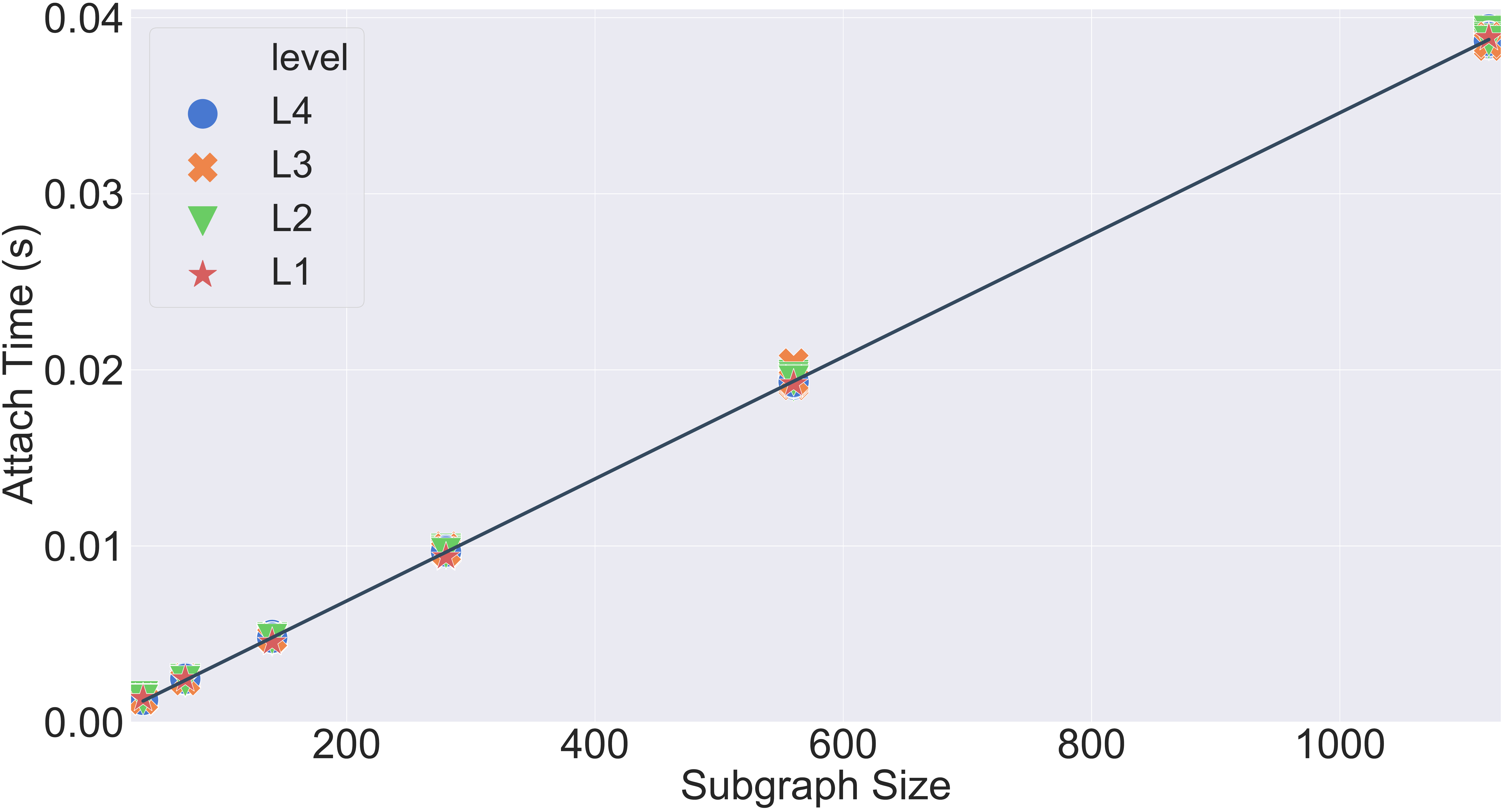}}
\caption{Subgraph add/update times by
subgraph request size.  The horizontal axis indicates the size of the
subgraph to be added, and the vertical axis indicates the time to add
the subgraph into each level's resource graph and then update the resource
graph's scheduling data. The regression models
are plotted as well to give an indication of the models' performance.}
\label{fig:attach-reg}
\end{figure}

In Section~\ref{subsubsec:add-update} we contended that the time 
to add a new subgraph and update the metadata in a level's local 
resource graph is linearly dependent on the number of vertices and 
edges ($n$) in the subgraph.  Here we adopt our strategy from the previous 
section and posit a linear model: $t_{add\_upd} = \beta n + \beta_0$. 
As before, we validate the linear regression with five-fold cross validation. 
Table~\ref{tab:reg} includes the average (across the five cross validation 
tests) MAPE and $R^2$ values.  The linear model exhibits a low error and 
high $R^2$ like the communication models, indicating that it 
represents the data with high fidelity and without omitting variables. 
Note that the intercept in Table~\ref{tab:reg} is exactly zero. 
The Scikit-learn regression model returns a small negative 
(thus unphysical) value, so we set the intercept to zero to 
obtain the coefficients in the table.  
Figure~\ref{fig:attach-reg} provides a visual representation 
of the excellent fit of the linear model to the data.  It also 
highlights the small spread and comparable medians of the 
measurements across different levels, providing a more 
comprehensive view of Figure~\ref{fig:attach-2240}.  

\subsection{Match model}
\label{subsec:match-model}

Figure~\ref{fig:attach-2240} in Section~\ref{subsubsec:add-update} demonstrates that 
the time to return a null match depends on the resource graph size, but not 
the request subgraph size for the jobspecs in the test.  In general, however, an arbitrarily complicated 
jobspec will require more time to return a null match than one with simple 
core and node requirements. Furthermore, for a request with matching resources, there 
is a complicated dependence on the request subgraph size and topology. The 
behavior is due to the relationship between the pruning filters used and 
the configured level of detail in the resource graph \cite{fluxion}.
Instead of embarking upon an in-depth study of the performance space of 
resource graph size and topology, and request graph size and topology 
(which is not our goal in this work), in this section we derive an upper 
bound for the time needed to find matching resources. Such an upper 
bound provides insight into the scalability of the nested \textproc{MatchGrow} 
procedure.

\begin{table*}[htbp]
\caption{Regression model cross validation results and coefficients to five significant digits}
\begin{center}
\begin{tabular}{|l|c|c|c|c|}
\hline
model &\multicolumn{2}{|c|}{CV results} & \multicolumn{2}{|c|}{reg coefs}\\
\cline{2-5}
& avg MAPE & avg $R^2$ & $\beta$ & $\beta_0$ \\
\hline
\textbf{L0 comm} & 0.0090208 & 0.99774 & $1.5829\times10^{-5}$ & 0.0020992 \\
\hline
\textbf{L1-4 comm} & 0.0027139 & 0.99990 & $9.0824\times10^{-6}$ & 0.00063196 \\
\hline
\textbf{attach} & 0.0088698 & 0.99991 & $3.4583\times10^{-5}$ & 0.0 \\
\hline
\end{tabular}
\label{tab:reg}
\end{center}
\end{table*}

Let $b$ be the minimum branching factor between levels in a nested Fluxion job represented by a 
balanced tree with $b > 1$. Let $s_0 = |V| + |E|$, where $G_0 = (V,E)$ is the top-level resource graph (level 0).
Assume the time for Fluxion to complete a \textproc{MatchGrow} for a \textit{fixed request size} 
on a single level is linear in the sum of graph vertices and edges $x$, i.e.: $\beta x + \beta_0$.

Now we want to derive a loose upper bound for the time to complete a match allocate for 
all levels.  To do so, we make a conservative assumption that the linear model can be 
applied at each level, regardless of whether the level finds a match (which it does not). 
In fact the time to return a null match is much less.
Then the time to complete a \textproc{MatchGrow} at level 0 is $t_0 = \beta s_0 + \beta_0 \implies t_0 > \beta s_0$ 
since $\beta_0 > 0$ (it must take positive time for a null resource graph). Then at some 
level $m$, $t_0b^{-m} > \beta s_0b^{-m}$ so the sum of the first term over all 
nested levels is:
\begin{equation}
\begin{split}
    \sum_{k=0}^{\log_{b}s_0-1} t_0b^{-k} &> \sum_{k=0}^{\log_{b}s_0-1} \beta s_0b^{-k} \\
\end{split}
\end{equation}
Where $\log_{b}s_0$ is the maximum number of levels for a graph of size and order $s_0$ with branching factor $b$. 
Now applying the Geometric Sum Formula to the left hand side yields:
\begin{equation}
    \begin{split}
        \sum_{k=0}^{\log_{b}s_0-1} t_0b^{-k} &= t_0\Big(\frac{1-b^{-\log_{b}s_0}}{1-b^{-1}}\Big) \\
        &= t_0b\Big(\frac{1-\frac{1}{s_0}}{b-1}\Big) \\
    \end{split}
\end{equation}
And adding the intercept term back in $\log_{b}s_0$ times gives an upper bound for the total time:
\begin{equation}
    \begin{split}
        t_0b\Big(\frac{1-\frac{1}{s_0}}{b-1}\Big) +  \beta_0\log_{b}s_0 > \sum_{k=0}^{\log_{b}s_0-1} \beta s_0b^{-k} + \beta_0 \\
    \end{split}
\end{equation}
Which for large $s_0$, $t_0 \gg \beta_0$, and $b=2$ (which is approximately the case 
in our experiments; see Section~\ref{subsec:nested-mg}) is $\approx 2t_0$. Note that 
$\beta_0$ is the time needed to return a null match for a null resource graph, so 
our assumption above is justified.  With such a bound in place, we just need to know 
how long it takes to perform \textproc{MatchGrow} at a single level with sufficient 
resources, which is nearly identical to a \textproc{MatchAllocate} 
(see Algorithm~\ref{algo:butd-grow-add}).

\subsection{Applying the model}
\label{subsec:gpu-app}

In the previous sections we develop components models for each of $t_{match} + t_{comms} + t_{add\_upd}$. 
The full model can be written (to three significant digits) as:
\begin{equation}
\begin{split}
   t_{MG} &= 2t_0 + m(1.58\times10^{-5}n + 0.00210) \\
          &+ p(9.08\times10^{-6}n + 0.000632) + qn3.46\times10^{-5} \\
\end{split}
\end{equation}
Where $n$ is the sum of vertices and edges in the subgraph request, $m$ is the number of parent-child 
levels communicating via IPoIB, $p$ is the number of local parent-child pairs, and $q$ is the number 
of nested levels. In our experiments from Section~\ref{sec:experiments} the values are: $m=1$, $p=3$, $q=4$.  
To evaluate our model, we use a jobspec requesting multiple node-local resources 
and compare our model's predicted time to the observed time.  Specifically, the jobspec 
requests one node with 4 GPUs and two sockets, each with 16 CPUs and 4GB memory.  The 
equivalent subgraph representation is size 94 vertices and edges.  Table~\ref{tab:model-preds} 
lists the component model and the corresponding MAPE value.  Since the overall 
model is linear, the total MAPE is just the average of the component errors. 

\begin{table}[htbp]
\caption{Component model prediction accuracy}
\begin{center}
\begin{tabular}{|c|c|}
\hline
\textbf{component model} & \textbf{MAPE} \\
\hline
\textbf{$t_{match}$} & 16.106 \\
\hline
\textbf{$t_{comms}$} & 0.0039155 \\
\hline
\textbf{$t_{add\_upd}$} & 0.0077445 \\
\hline
\end{tabular}
\label{tab:model-preds}
\end{center}
\end{table}

The MAPE values in  Table~\ref{tab:model-preds} for the $t_{comms}$ and $t_{add\_upd}$ component models are 
very low, demonstrating that the models generalize well to a new (and more complex) 
resource subgraph request. The MAPE value for the $t_{match}$ model is moderate. 
The loose upper bound derived in Section~\ref{subsec:match-model} contains several 
assumptions that guarantee the bound but contribute to its looseness. First, the 
worst-case assumption that there are $\log_{b}s_0$ levels translates to 
14 levels (for our resource graph of size 18,061) rather than the five 
present in our tests.  
Ultimately $t_{match}$ is dominated by the single successful \textproc{MatchAllocate} 
(unsuccessful calls are much faster) which is approximately the same scenario as a single-level \textproc{MatchAllocate}. 
As Fluxion has proved to be fast and scalable for 
single-level \textproc{MatchAllocate} \cite{flux-fgcs:2020} 
we conclude that it is also fast and scalable for nested \textproc{MatchGrow}.


\section{Related Work}
\label{sec:related-work}

In Prabhakaran et al, 2014 \cite{prabhakaran2014} the group 
proposes a system for dynamically allocating resources to 
evolving applications. The system balances fairness between 
rigid and evolving jobs, and is implemented in Maui.  
Prabhakaran et al, 2018 \cite{prabhakaran2018} create  
a method for dynamic replacement of failed nodes. 
To do so the method uses Maui to allocate nodes from those in use 
by malleable applications, or allocates free nodes from a 
pool of spares. 
The Slurm resource manager \cite{yoo2003} supports expand and shrink operations by allowing a
running job to submit a new job with a dependency indicator and then merging the
allocations. The approach used by Slurm demands
that a job with dynamic resources release all the nodes
that came with a single dynamic request. 
The works by Prabhakaran et al and the Slurm RJMS 
differ from our study in that the total number 
of resources does not change; the jobs are elastic, 
but the clusters are not.  Our work enables both forms 
of dynamism.

Marshall et al, \cite{marshall2010} develop an ``elastic-site''
model that enables on-premise services to use elastically-provisioned 
cloud resources.  Their model is built on the Nimbus Platform.
The main focus of the work is to determine 
when to burst, rather than how; to that extent 
the work complements ours.  Netto et al, 2018 \cite{netto2018} provide a comprehensive 
survey of the investigations and remaining challenges 
related to running HPC applications in the cloud. 
The suitability of applications for cloud execution 
is explored in several studies.  In particular, 
Clemente-Castelló et al \cite{clemente2017} derive a cost model for cloud bursting 
MapReduce workflows.



\section{Conclusions}
\label{sec:discuss-conclusions}
%

Emerging complex scientific workflows demand extreme, 
dynamic resource heterogeneity supplied by the latest 
HPC and cloud converged systems. The massive complexity 
of the new systems translates to an explosive growth 
of the resource configuration space, which is 
becoming unmanageable by current HPC RJMS. 
We identify three key capabilities needed to 
manage the configuration space effectively: 
RJMS elasticity to accommodate elastic jobs, 
specialization of external resources to manage 
resources selected by the resource provider, 
and the capability to schedule cloud orchestration framework tasks.
These capabilities have been studied individually, but 
to the best of our knowledge our work is the first 
to consider and address all three core capabilities combined. 
We propose a novel way to provide 
the capabilities via integrating a dynamic, 
directed graph-based resource model with fully hierarchical scheduling, and 
define procedures to edit the resource graph 
efficiently. We embody the capabilities in 
Fluxion, a new graph-based RJMS, by 
imbuing it with resource dynamism and 
bursting capability. We then run a series of 
experiments to evaluate the suitability of our solution
for managing massive resource configuration spaces.
Our results demonstrate that our proposed solution
can provide the three key capabilities 
for managing extreme, dynamic heterogeneity. Our solution lights a beacon 
for next-generation computing centers that will use abundant elasticity
in dynamic workflows and convergence with the cloud.

\section*{Acknowledgments}

We are grateful to the entire Flux team  
for their help addressing test-related issues with 
software and configurations. In particular, Jim Garlick 
and Mark Grondona provided invaluable advice with 
experimental design and optimization, as well as 
help with diagnosing implementation problems. 
We wish to thank Jeff Buseman, David Fox, and Robin 
Goldstone for creating and configuring our test cluster 
and for enabling our AWS EC2 tests.

This work was performed under the auspices of the U.S. Department of Energy by
Lawrence Livermore National Laboratory under Contract DE-AC52-07NA27344
(LLNL-JRNL-826450-DRAFT).

\bibliographystyle{ACM-Reference-Format}
\bibliography{arxiv-2021.bib}


\begin{thebibliography}{26}


\ifx \showCODEN    \undefined \def \showCODEN     #1{\unskip}     \fi
\ifx \showDOI      \undefined \def \showDOI       #1{#1}\fi
\ifx \showISBNx    \undefined \def \showISBNx     #1{\unskip}     \fi
\ifx \showISBNxiii \undefined \def \showISBNxiii  #1{\unskip}     \fi
\ifx \showISSN     \undefined \def \showISSN      #1{\unskip}     \fi
\ifx \showLCCN     \undefined \def \showLCCN      #1{\unskip}     \fi
\ifx \shownote     \undefined \def \shownote      #1{#1}          \fi
\ifx \showarticletitle \undefined \def \showarticletitle #1{#1}   \fi
\ifx \showURL      \undefined \def \showURL       {\relax}        \fi
\providecommand\bibfield[2]{#2}
\providecommand\bibinfo[2]{#2}
\providecommand\natexlab[1]{#1}
\providecommand\showeprint[2][]{arXiv:#2}

\bibitem[\protect\citeauthoryear{??}{bgl}{2002}]%
        {bgl2002}
 \bibinfo{year}{2002}\natexlab{}.
\newblock \bibinfo{booktitle}{\emph{The Boost Graph Library: User Guide and
  Reference Manual}}.
\newblock \bibinfo{publisher}{Addison-Wesley Longman Publishing Co., Inc.},
  \bibinfo{address}{USA}.
\newblock
\showISBNx{0201729148}


\bibitem[\protect\citeauthoryear{??}{lsf}{2021}]%
        {lsf}
 \bibinfo{year}{2021}\natexlab{}.
\newblock \bibinfo{title}{{IBM} {Spectrum} {LSF} resource connector}.
\newblock
  \bibinfo{howpublished}{\url{https://www.ibm.com/support/knowledgecenter/SSWRJV_10.1.0/lsf_welcome/lsf_kc_resource_connector.html}}.
\newblock
\newblock
\shownote{Retrieved February 10, 2021.}


\bibitem[\protect\citeauthoryear{??}{psc}{2021}]%
        {psc-bridges-2}
 \bibinfo{year}{2021}\natexlab{}.
\newblock \bibinfo{title}{{Pittsburgh Supercomputing Center Bridges-2
  Supercomputer}}.
\newblock
  \bibinfo{howpublished}{\url{https://www.psc.edu/resources/bridges-2/}}.
\newblock
\newblock
\shownote{Retrieved February 10, 2021.}


\bibitem[\protect\citeauthoryear{??}{sds}{2021}]%
        {sdsc-expanse}
 \bibinfo{year}{2021}\natexlab{}.
\newblock \bibinfo{title}{{San Diego Supercomputer Center Expanse
  Supercomputer}}.
\newblock
  \bibinfo{howpublished}{\url{https://www.sdsc.edu/services/hpc/expanse/}}.
\newblock
\newblock
\shownote{Retrieved February 10, 2021.}


\bibitem[\protect\citeauthoryear{??}{slu}{2021}]%
        {slurm-csg}
 \bibinfo{year}{2021}\natexlab{}.
\newblock \bibinfo{title}{{SchedMD Cloud Scheduling Guide}}.
\newblock
  \bibinfo{howpublished}{\url{https://slurm.schedmd.com/elastic_computing.html}}.
\newblock
\newblock
\shownote{Retrieved February 10, 2021.}


\bibitem[\protect\citeauthoryear{??}{vol}{2021}]%
        {volcano}
 \bibinfo{year}{2021}\natexlab{}.
\newblock \bibinfo{title}{{Volcano Kubernetes native batch system}}.
\newblock \bibinfo{howpublished}{\url{https://volcano.sh/en/}}.
\newblock
\newblock
\shownote{Retrieved February 10, 2021.}


\bibitem[\protect\citeauthoryear{Ahn et~al\mbox{.}}{Ahn et~al\mbox{.}}{2020}]%
        {flux-fgcs:2020}
\bibfield{author}{\bibinfo{person}{Dong~H. Ahn} {et~al\mbox{.}}}
  \bibinfo{year}{2020}\natexlab{}.
\newblock \showarticletitle{Flux: Overcoming scheduling challenges for exascale
  workflows}.
\newblock \bibinfo{journal}{\emph{Future Generation Computer Systems}}
  \bibinfo{volume}{110} (\bibinfo{year}{2020}), \bibinfo{pages}{202 -- 213}.
\newblock
\showISSN{0167-739X}


\bibitem[\protect\citeauthoryear{Altair Engineering Inc.}{Altair Engineering
  Inc.}{2020}]%
        {pbspro}
Altair Engineering Inc. \bibinfo{year}{2020}\natexlab{}.
\newblock \bibinfo{booktitle}{\emph{{Altair PBS Professional 2020.1 Cloud
  Guide}}}.
\newblock Altair Engineering Inc.
\newblock


\bibitem[\protect\citeauthoryear{{Clemente-Castelló}, {Mayo}, and
  {Fernández}}{{Clemente-Castelló} et~al\mbox{.}}{2017}]%
        {clemente2017}
\bibfield{author}{\bibinfo{person}{F.~J. {Clemente-Castelló}},
  \bibinfo{person}{R. {Mayo}}, {and} \bibinfo{person}{J.~C. {Fernández}}.}
  \bibinfo{year}{2017}\natexlab{}.
\newblock \showarticletitle{{Cost Model and Analysis of Iterative MapReduce
  Applications for Hybrid Cloud Bursting}}. In \bibinfo{booktitle}{\emph{2017
  17th IEEE/ACM International Symposium on Cluster, Cloud and Grid Computing
  (CCGRID)}}. \bibinfo{pages}{858--864}.
\newblock


\bibitem[\protect\citeauthoryear{Di~Natale et~al\mbox{.}}{Di~Natale
  et~al\mbox{.}}{2019}]%
        {dinatale2019}
\bibfield{author}{\bibinfo{person}{Francesco Di~Natale} {et~al\mbox{.}}}
  \bibinfo{year}{2019}\natexlab{}.
\newblock \showarticletitle{{A Massively Parallel Infrastructure for Adaptive
  Multiscale Simulations: Modeling RAS Initiation Pathway for Cancer}}. In
  \bibinfo{booktitle}{\emph{Proceedings of the International Conference for
  High Performance Computing, Networking, Storage and Analysis}} (Denver,
  Colorado) \emph{(\bibinfo{series}{SC '19})}. \bibinfo{publisher}{Association
  for Computing Machinery}, \bibinfo{address}{New York, NY, USA}, Article
  \bibinfo{articleno}{57}, \bibinfo{numpages}{16}~pages.
\newblock
\showISBNx{9781450362290}


\bibitem[\protect\citeauthoryear{Feitelson and Rudolph}{Feitelson and
  Rudolph}{1996}]%
        {feitelson1996}
\bibfield{author}{\bibinfo{person}{Dror~G. Feitelson} {and}
  \bibinfo{person}{Larry Rudolph}.} \bibinfo{year}{1996}\natexlab{}.
\newblock \showarticletitle{{Towards Convergence in Job Schedulers for Parallel
  Supercomputers}}. In \bibinfo{booktitle}{\emph{Proceedings of the Workshop on
  Job Scheduling Strategies for Parallel Processing}}
  \emph{(\bibinfo{series}{IPPS ’96})}. \bibinfo{publisher}{Springer-Verlag},
  \bibinfo{address}{Berlin, Heidelberg}, \bibinfo{pages}{1–26}.
\newblock
\showISBNx{3540618643}


\bibitem[\protect\citeauthoryear{{Gupta}, {Acun}, {Sarood}, and
  {Kalé}}{{Gupta} et~al\mbox{.}}{2014}]%
        {gupta2014}
\bibfield{author}{\bibinfo{person}{A. {Gupta}}, \bibinfo{person}{B. {Acun}},
  \bibinfo{person}{O. {Sarood}}, {and} \bibinfo{person}{L.~V. {Kalé}}.}
  \bibinfo{year}{2014}\natexlab{}.
\newblock \showarticletitle{Towards realizing the potential of malleable jobs}.
  In \bibinfo{booktitle}{\emph{2014 21st International Conference on High
  Performance Computing (HiPC)}}. \bibinfo{pages}{1--10}.
\newblock


\bibitem[\protect\citeauthoryear{{Gupta}, {Kale}, {Gioachin}, {March}, {Suen},
  {Lee}, {Faraboschi}, {Kaufmann}, and {Milojicic}}{{Gupta}
  et~al\mbox{.}}{2013}]%
        {gupta2013}
\bibfield{author}{\bibinfo{person}{A. {Gupta}}, \bibinfo{person}{L.~V. {Kale}},
  \bibinfo{person}{F. {Gioachin}}, \bibinfo{person}{V. {March}},
  \bibinfo{person}{C.~H. {Suen}}, \bibinfo{person}{B. {Lee}},
  \bibinfo{person}{P. {Faraboschi}}, \bibinfo{person}{R. {Kaufmann}}, {and}
  \bibinfo{person}{D. {Milojicic}}.} \bibinfo{year}{2013}\natexlab{}.
\newblock \showarticletitle{{The Who, What, Why, and How of High Performance
  Computing in the Cloud}}. In \bibinfo{booktitle}{\emph{2013 IEEE 5th
  International Conference on Cloud Computing Technology and Science}},
  Vol.~\bibinfo{volume}{1}. \bibinfo{pages}{306--314}.
\newblock


\bibitem[\protect\citeauthoryear{{Kale}, {Kumar}, and {DeSouza}}{{Kale}
  et~al\mbox{.}}{2002}]%
        {kale2002}
\bibfield{author}{\bibinfo{person}{L.~V. {Kale}}, \bibinfo{person}{S. {Kumar}},
  {and} \bibinfo{person}{J. {DeSouza}}.} \bibinfo{year}{2002}\natexlab{}.
\newblock \showarticletitle{{A Malleable-Job System for Timeshared Parallel
  Machines}}. In \bibinfo{booktitle}{\emph{2nd IEEE/ACM International Symposium
  on Cluster Computing and the Grid (CCGRID'02)}}. \bibinfo{pages}{230--230}.
\newblock


\bibitem[\protect\citeauthoryear{{Lawrence Livermore National
  Laboratory}}{{Lawrence Livermore National Laboratory}}{2014}]%
        {fluxion}
\bibfield{author}{\bibinfo{person}{{Lawrence Livermore National Laboratory}}.}
  \bibinfo{year}{2014}\natexlab{}.
\newblock \bibinfo{title}{Fluxion}.
\newblock
  \bibinfo{howpublished}{\url{https://github.com/flux-framework/flux-sched}}.
\newblock


\bibitem[\protect\citeauthoryear{{Marshall}, {Keahey}, and
  {Freeman}}{{Marshall} et~al\mbox{.}}{2010}]%
        {marshall2010}
\bibfield{author}{\bibinfo{person}{P. {Marshall}}, \bibinfo{person}{K.
  {Keahey}}, {and} \bibinfo{person}{T. {Freeman}}.}
  \bibinfo{year}{2010}\natexlab{}.
\newblock \showarticletitle{{Elastic Site: Using Clouds to Elastically Extend
  Site Resources}}. In \bibinfo{booktitle}{\emph{2010 10th IEEE/ACM
  International Conference on Cluster, Cloud and Grid Computing}}.
  \bibinfo{pages}{43--52}.
\newblock


\bibitem[\protect\citeauthoryear{Minnich et~al\mbox{.}}{Minnich
  et~al\mbox{.}}{2020}]%
        {minnich2020}
\bibfield{author}{\bibinfo{person}{Amanda~J. Minnich} {et~al\mbox{.}}}
  \bibinfo{year}{2020}\natexlab{}.
\newblock \showarticletitle{{AMPL: A Data-Driven Modeling Pipeline for Drug
  Discovery}}.
\newblock \bibinfo{journal}{\emph{Journal of Chemical Information and
  Modeling}} \bibinfo{volume}{60}, \bibinfo{number}{4} (\bibinfo{year}{2020}),
  \bibinfo{pages}{1955--1968}.
\newblock
\newblock
\shownote{PMID: 32243153.}


\bibitem[\protect\citeauthoryear{Netto, Calheiros, Rodrigues, Cunha, and
  Buyya}{Netto et~al\mbox{.}}{2018}]%
        {netto2018}
\bibfield{author}{\bibinfo{person}{Marco A.~S. Netto},
  \bibinfo{person}{Rodrigo~N. Calheiros}, \bibinfo{person}{Eduardo~R.
  Rodrigues}, \bibinfo{person}{Renato L.~F. Cunha}, {and}
  \bibinfo{person}{Rajkumar Buyya}.} \bibinfo{year}{2018}\natexlab{}.
\newblock \showarticletitle{{HPC Cloud for Scientific and Business
  Applications: Taxonomy, Vision, and Research Challenges}}.
\newblock \bibinfo{journal}{\emph{ACM Comput. Surv.}} \bibinfo{volume}{51},
  \bibinfo{number}{1}, Article \bibinfo{articleno}{8} (\bibinfo{date}{Jan.}
  \bibinfo{year}{2018}), \bibinfo{numpages}{29}~pages.
\newblock
\showISSN{0360-0300}


\bibitem[\protect\citeauthoryear{Pedregosa, Varoquaux, Gramfort, Michel,
  Thirion, Grisel, Blondel, Prettenhofer, Weiss, Dubourg, Vanderplas, Passos,
  Cournapeau, Brucher, Perrot, and Duchesnay}{Pedregosa et~al\mbox{.}}{2011}]%
        {scikit-learn2011}
\bibfield{author}{\bibinfo{person}{F. Pedregosa}, \bibinfo{person}{G.
  Varoquaux}, \bibinfo{person}{A. Gramfort}, \bibinfo{person}{V. Michel},
  \bibinfo{person}{B. Thirion}, \bibinfo{person}{O. Grisel},
  \bibinfo{person}{M. Blondel}, \bibinfo{person}{P. Prettenhofer},
  \bibinfo{person}{R. Weiss}, \bibinfo{person}{V. Dubourg}, \bibinfo{person}{J.
  Vanderplas}, \bibinfo{person}{A. Passos}, \bibinfo{person}{D. Cournapeau},
  \bibinfo{person}{M. Brucher}, \bibinfo{person}{M. Perrot}, {and}
  \bibinfo{person}{E. Duchesnay}.} \bibinfo{year}{2011}\natexlab{}.
\newblock \showarticletitle{{Scikit-learn: Machine Learning in {P}ython}}.
\newblock \bibinfo{journal}{\emph{Journal of Machine Learning Research}}
  \bibinfo{volume}{12} (\bibinfo{year}{2011}), \bibinfo{pages}{2825--2830}.
\newblock


\bibitem[\protect\citeauthoryear{{Prabhakaran}, {Iqbal}, {Rinke}, {Windisch},
  and {Wolf}}{{Prabhakaran} et~al\mbox{.}}{2014}]%
        {prabhakaran2014}
\bibfield{author}{\bibinfo{person}{S. {Prabhakaran}}, \bibinfo{person}{M.
  {Iqbal}}, \bibinfo{person}{S. {Rinke}}, \bibinfo{person}{C. {Windisch}},
  {and} \bibinfo{person}{F. {Wolf}}.} \bibinfo{year}{2014}\natexlab{}.
\newblock \showarticletitle{{A Batch System with Fair Scheduling for Evolving
  Applications}}. In \bibinfo{booktitle}{\emph{2014 43rd International
  Conference on Parallel Processing}}. \bibinfo{pages}{351--360}.
\newblock


\bibitem[\protect\citeauthoryear{{Prabhakaran}, {Neumann}, {Rinke}, {Wolf},
  {Gupta}, and {Kale}}{{Prabhakaran} et~al\mbox{.}}{2015}]%
        {prabhakaran2015}
\bibfield{author}{\bibinfo{person}{S. {Prabhakaran}}, \bibinfo{person}{M.
  {Neumann}}, \bibinfo{person}{S. {Rinke}}, \bibinfo{person}{F. {Wolf}},
  \bibinfo{person}{A. {Gupta}}, {and} \bibinfo{person}{L.~V. {Kale}}.}
  \bibinfo{year}{2015}\natexlab{}.
\newblock \showarticletitle{{A Batch System with Efficient Adaptive Scheduling
  for Malleable and Evolving Applications}}. In \bibinfo{booktitle}{\emph{2015
  IEEE International Parallel and Distributed Processing Symposium}}.
  \bibinfo{pages}{429--438}.
\newblock


\bibitem[\protect\citeauthoryear{{Prabhakaran}, {Neumann}, and
  {Wolf}}{{Prabhakaran} et~al\mbox{.}}{2018}]%
        {prabhakaran2018}
\bibfield{author}{\bibinfo{person}{S. {Prabhakaran}}, \bibinfo{person}{M.
  {Neumann}}, {and} \bibinfo{person}{F. {Wolf}}.}
  \bibinfo{year}{2018}\natexlab{}.
\newblock \showarticletitle{Efficient Fault Tolerance Through Dynamic Node
  Replacement}. In \bibinfo{booktitle}{\emph{2018 18th IEEE/ACM International
  Symposium on Cluster, Cloud and Grid Computing (CCGRID)}}.
  \bibinfo{pages}{163--172}.
\newblock


\bibitem[\protect\citeauthoryear{Stanzione, West, Evans, Minyard, Ghattas, and
  Panda}{Stanzione et~al\mbox{.}}{2020}]%
        {stanzione2020}
\bibfield{author}{\bibinfo{person}{Dan Stanzione}, \bibinfo{person}{John West},
  \bibinfo{person}{R.~Todd Evans}, \bibinfo{person}{Tommy Minyard},
  \bibinfo{person}{Omar Ghattas}, {and} \bibinfo{person}{Dhabaleswar~K.
  Panda}.} \bibinfo{year}{2020}\natexlab{}.
\newblock \showarticletitle{{Frontera: The Evolution of Leadership Computing at
  the National Science Foundation}}. In \bibinfo{booktitle}{\emph{Practice and
  Experience in Advanced Research Computing}} (Portland, OR, USA)
  \emph{(\bibinfo{series}{PEARC '20})}. \bibinfo{publisher}{Association for
  Computing Machinery}, \bibinfo{address}{New York, NY, USA},
  \bibinfo{pages}{106–111}.
\newblock
\showISBNx{9781450366892}


\bibitem[\protect\citeauthoryear{Vetter et~al\mbox{.}}{Vetter
  et~al\mbox{.}}{2018}]%
        {ascr2018}
\bibfield{author}{\bibinfo{person}{Jeffrey~S. Vetter} {et~al\mbox{.}}}
  \bibinfo{year}{2018}\natexlab{}.
\newblock \showarticletitle{{Extreme Heterogeneity 2018 - Productive
  Computational Science in the Era of Extreme Heterogeneity: Report for DOE
  ASCR Workshop on Extreme Heterogeneity}}.
\newblock  (\bibinfo{date}{12} \bibinfo{year}{2018}).
\newblock


\bibitem[\protect\citeauthoryear{Yang, Domeniconi, Zhang, and Cong}{Yang
  et~al\mbox{.}}{2020}]%
        {yang2020}
\bibfield{author}{\bibinfo{person}{Chih-Chieh Yang}, \bibinfo{person}{Giacomo
  Domeniconi}, \bibinfo{person}{Leili Zhang}, {and} \bibinfo{person}{Guojing
  Cong}.} \bibinfo{year}{2020}\natexlab{}.
\newblock \showarticletitle{{Design of AI-Enhanced Drug Lead Optimization
  Workflow for HPC and Cloud}}. In \bibinfo{booktitle}{\emph{IEEE International
  Conference on Big Data}}. \bibinfo{pages}{5861--5863}.
\newblock


\bibitem[\protect\citeauthoryear{Yoo, Jette, and Grondona}{Yoo
  et~al\mbox{.}}{2003}]%
        {yoo2003}
\bibfield{author}{\bibinfo{person}{Andy~B. Yoo}, \bibinfo{person}{Morris~A.
  Jette}, {and} \bibinfo{person}{Mark Grondona}.}
  \bibinfo{year}{2003}\natexlab{}.
\newblock \showarticletitle{{SLURM: Simple Linux Utility for Resource
  Management}}. In \bibinfo{booktitle}{\emph{Job Scheduling Strategies for
  Parallel Processing}}, \bibfield{editor}{\bibinfo{person}{Dror Feitelson},
  \bibinfo{person}{Larry Rudolph}, {and} \bibinfo{person}{Uwe Schwiegelshohn}}
  (Eds.). \bibinfo{publisher}{Springer Berlin Heidelberg},
  \bibinfo{address}{Berlin, Heidelberg}, \bibinfo{pages}{44–60}.
\newblock
\showISBNx{978-3-540-39727-4}


\end{thebibliography}

\end{document}